\documentclass{article}
\usepackage{alife9}
\usepackage{graphicx}

\begin{document}

\title{Updating Schemes in Random Boolean Networks: Do They Really Matter?}
\author{Carlos Gershenson \\
\mbox{} Centrum Leo Apostel, Vrije Universiteit Brussel. Krijgskundestraat
33 B-1160 Brussel, Belgium\\
cgershen@vub.ac.be \ http://homepages.vub.ac.be/\symbol{126}cgershen}
\maketitle

\begin{abstract}
In this paper we try to end the debate concerning the suitability of
different updating schemes in random Boolean networks (RBNs). We quantify
for the first time loose attractors in asyncrhonous RBNs, which allows us to
analyze the \emph{complexity reduction} related to different updating
schemes. We also report that all updating schemes yield very similar
critical stability values, meaning that the ``edge of chaos'' does not
depend much on the updating scheme. After discussion, we conclude that
synchonous RBNs are justifiable theoretical models of biological networks.
\end{abstract}

\section{Introduction}

Random Boolean Networks (RBNs) have been useful tools for modelling and
studying the functional and computational requirements and possibilities of
life \cite{Kauffman1993}. They are simple and general. Their advantage is
that one does not need to assume any previous functionality. Exploring
different parameters of the network, such as number of nodes and
connections, one can find regions in the parameter space where computation,
such as the one required by life, is very probable. This region is
characterized by being stable enough to keep information, but flexible
enough to transmit, manipulate and transform it. Not too frozen and ordered,
but not too variable and chaotic. That is why it has been referred to as the 
\emph{edge of chaos}.

There has been a debate on wether RBNs should have a synchronous or
asynchronous, deterministic or non-deterministic, updating scheme \cite
{HarveyBossomaier1997,DiPaolo2001,Gershenson2002e,GershensonEtAl2003a}. We
attempt to close this debate in this paper motivated by the results
presented within.

In the next section, we present the background of RBNs and different
updating schemes. Then, we show results of experiments which include the
quantification of loose attractors, and also present results related to the
stability of different RBNs. The main discussion follows from these results.
Concluding remarks close the paper.

\section{Background}

RBNs are a generalization of cellular automata (CA), where the functionality
of each node is not restricted to its neighbourhood. The state (zero or one)
of the $n$ nodes of a RBN depends on the states of $k$ other nodes
connecting to each. Which nodes affect which (the connectivity) is initially
generated at random. The way nodes affect each other (the functionality) is
determined by logic lookup tables, which are also initially randomly
generated. We can bias the connectivity to achieve different topologies,
such as scale-free \cite{Aldana2003}. We can bias the functionality as well,
and this will result in different network properties \cite{DerridaPomeau1986}%
. The RBNs we study have homogeneous topology and no functional bias, as
this is the standard in the literature.

We have proposed a classification of RBNs according to their updating scheme 
\cite{Gershenson2002e}. We have seen that the updating changes considerably
the properties of the same networks, such as number of attractors and
attractor lengths, and also affects drastically the shapes of the basins of
attraction. The change of updating scheme also affects drastically the
behaviour of models based on RBNs or CA \cite{BersiniDetours1994}.

Classical RBNs (\textbf{CRBNs}) \cite{Kauffman1969} have synchronous
updating:\ all nodes at time $t+1$ take into account nodes at time $t$ for
their updating. Since the dynamics are deterministic, and the state space is
finite, sooner or later a state will be repeated, and the network will have
reached an \emph{attractor}. If it consists of only one state, it will be a 
\emph{point} attractor. Otherwise, it will be a \emph{cycle} attractor.
CRBNs have been widely studied \cite{Wuensche1997,AldanaEtAl2003}.

Asynchronous RBNs (\textbf{ARBNs}) \cite{HarveyBossomaier1997} have
asynchronous and non-deterministic updating. A node is randomly chosen and
the network updated. There are point attractors, but no cycle attractors due
to the non-determinism. However, not all states are revisited. We can
identify as \emph{loose} attractors the regions of the state space which
capture the dynamics of the network. Until now, loose attractors had been
ignored in statistical studies, including ours.

Generalized Asynchronous RBNs (\textbf{GARBNs}) \cite{Gershenson2002e} are
similar to ARBNs but semi-synchronous: each time step they select randomly
which nodes to update synchronously. Therefore, at a time step some of the
nodes can be updated, only one, or even all of them.

Deterministic Asynchronous RBNs (\textbf{DARBNs}) \cite{Gershenson2002e}
introduce two parameters per node, $p$ and $q$, which allow asynchrony in a
deterministic fashion. A node is updated if the modulus of $p$ over time
equals $q$. The set of all $p$'s and $q$'s can be seen as the \emph{context}
of the network\cite{GershensonEtAl2003a}. The context is initially randomly
generated, where $p$'s can take integer values between one and $maxP$, and $%
q $'s between zero and $maxP-1$. DARBNs have point and cycle attractors.

Deterministic Generalized Asynchronous RBNs (\textbf{DGARBNs}) \cite
{Gershenson2002e} are the semi-synchronous coutnerpart of DARBNs:\ they
update synchronously all nodes which fulfill the condition $p\ mod\ t==q$.

Mixed-context RBNs (\textbf{MxRBNs}) \cite{GershensonEtAl2003a} are
non-deterministic in a particular way:\ They are DGARBNs with $m$ ``\emph{%
pure}'' contexts (sets of $p$'s and $q$'s), and each $P$ time steps, one
context is chosen randomly.

We have recently studied the sensitivity to initial conditions of different
types of RBNs \cite{GershensonSubmitted}, and we found out that the updating
scheme almost does not affect the phase transition between ``ordered'' and
``chaotic'' regimes of the networks.

\section{Experiments}

For our experiments we used a software laboratory we have been developing
for this purpose. It is an open source project, available at
http://rbn.sourceforge.net.

\subsection{Loose attractors}

An attractor can be seen as a part of the state space which a dynamical
system has a high probability of reaching\footnote{%
In a system with random state transitions, a loose ``attractor'' would be
equal to the whole state space. As we will expose, this is not at all the
case of non-deterministic RBNs: the set of their ``preferred'' states is
significantly smaller than the state space.}. In deterministic systems this
can be precisely defined, since from one state the dynamics will follow to
only one another state. However, in non-deterministic systems several states
can follow from one state. The state transitions can be analysed carefully
for specific systems, but this becomes intractable for large families of
networks. Therefore, we can approximate with simulations loose attractors in
RBNs as the part of the state space which is reached after some time.

Here we consider for the first time in statistical analysis the existence of
loose attractors. We believe that their study is very important, since they
reflect the complexity reduction which non-deterministic RBNs can achieve.
The algorithm we devised to find them is as follows: We first let run the
network from an initial state for it to reach an attractor (we used more
than 10000 time steps). Then, if the state is already in an attractor we
have previously found, we begin again with a different initial state. If
not, we create a new attractor composed of that state, and the states which
we obtain following the dynamic of the network are added to the attractor
(omitting states which were already visited). This search for states in an
attractor lasts until a maximum search period is reached (we used 3000
iterations). After this, another initial state is chosen and the algorithm
is run again. After all the possible initial states of a network, or a
certain number, have been explored, there might be some overlapping
attractors, since the state of a loose attractor might be missed at first by
the algorithm, but then it would construct a redundant attractor over it.
Therefore, a trimming process takes place, in which repeated states are
sought in different attractors, and if found, the attractors are merged,
removing redundant states.

The algorithm is general enough so that any type of attractor (point, cycle,
loose) can be detected with it, though it is redundant and computationally
expensive. The trimming of the attractors is necessary only for
non-deterministic networks (ARBN, GARBN, and MxRBN). Notice that for these
networks the sequence of states in the found attractors does not indicate
necessarily the actual possible state sequences. The algorithm cannot handle
networks of $n>20$ in a computer with 1Gb RAM. The best supercomputer now
would not achieve $n=30$. Probably the algorithm could be optimized. Also,
an analytical solution would be very helpful to explore loose attractors
further, but this is not an easy task.

\begin{figure}[t]
\begin{center}
\includegraphics[width=3.2in]{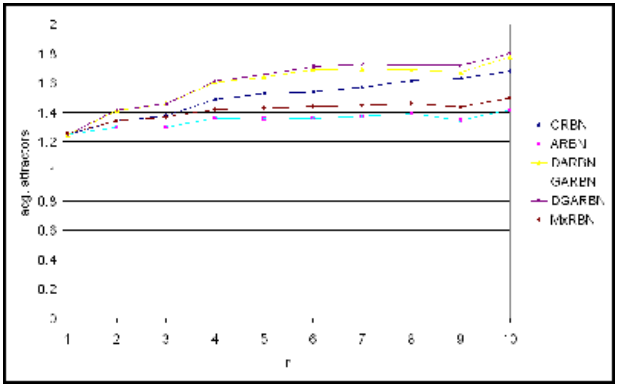}
\end{center}
\caption{Average number of attractors for k=1}
\label{attk1}
\end{figure}

\begin{figure}[t]
\begin{center}
\includegraphics[width=3.2in]{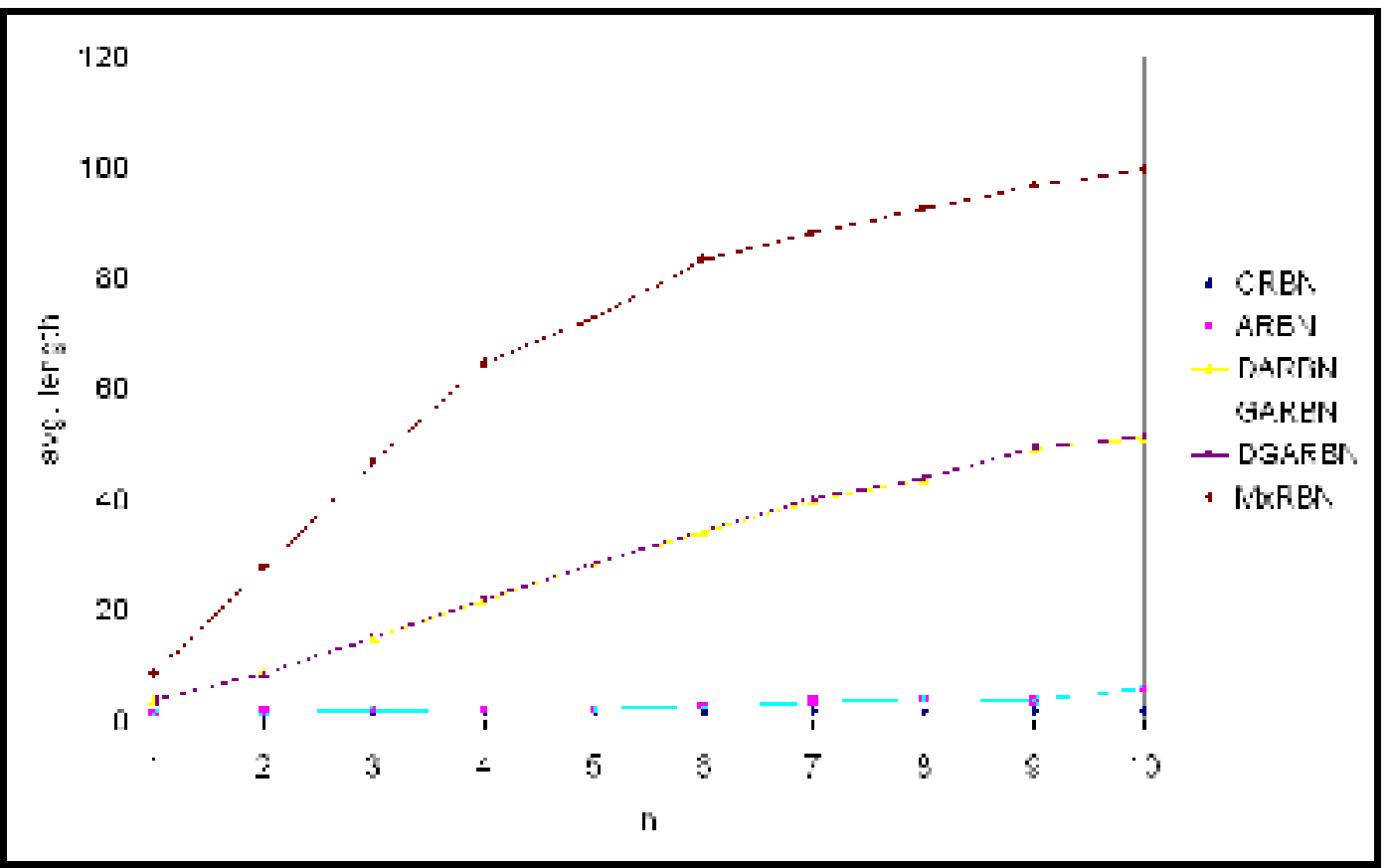}
\end{center}
\caption{Average length of attractors for k=1}
\label{lk1}
\end{figure}

\begin{figure}[t]
\begin{center}
\includegraphics[width=3.2in]{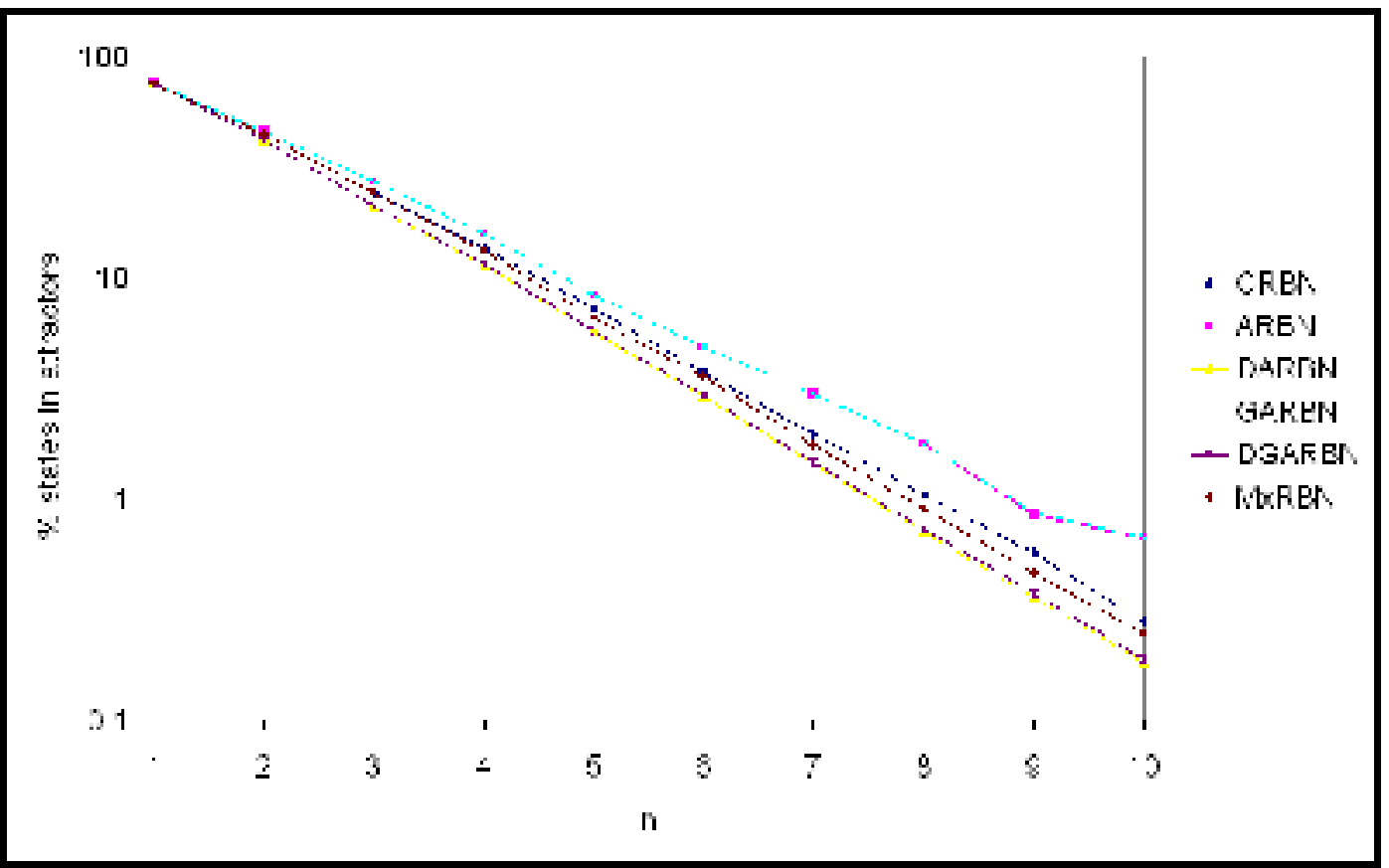}
\end{center}
\caption{Average percentage of states in attractors for k=1}
\label{pctk1}
\end{figure}

\begin{figure}[t]
\begin{center}
\includegraphics[width=3.2in]{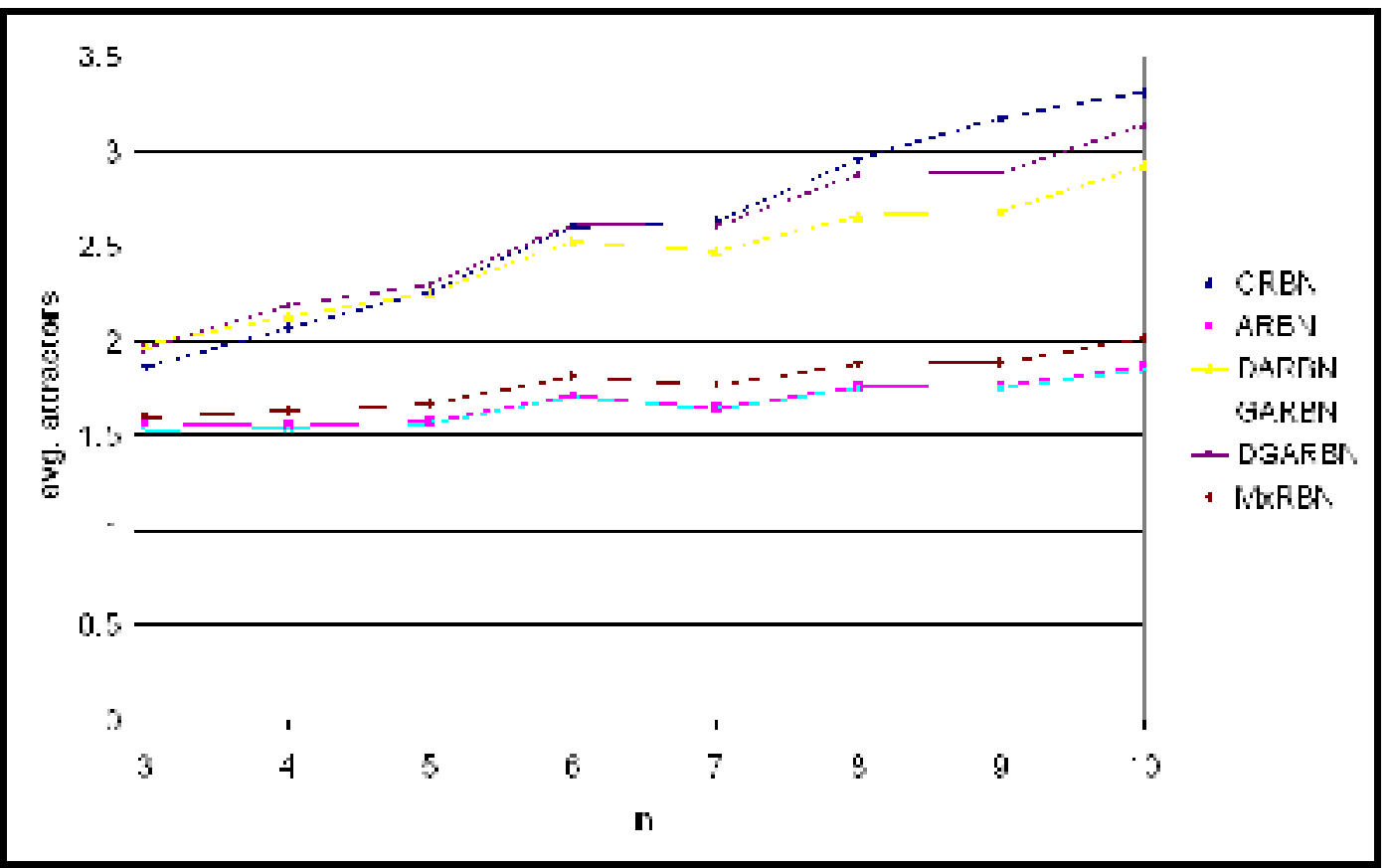}
\end{center}
\caption{Average number of attractors for k=3}
\label{attk3}
\end{figure}

\begin{figure}[t]
\begin{center}
\includegraphics[width=3.2in]{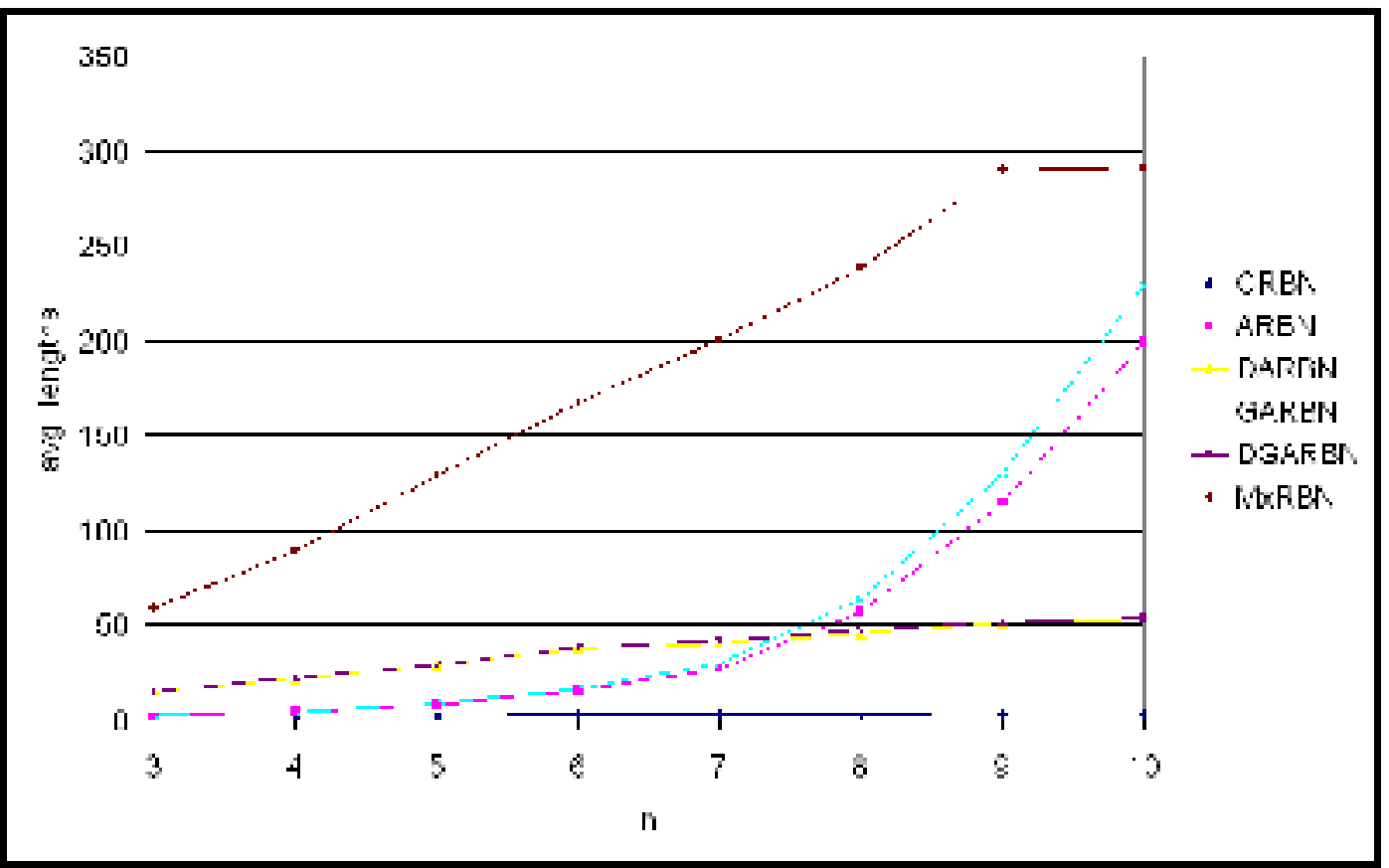}
\end{center}
\caption{Average length of attractors for k=3}
\label{lk3}
\end{figure}

\begin{figure}[t]
\begin{center}
\includegraphics[width=3.2in]{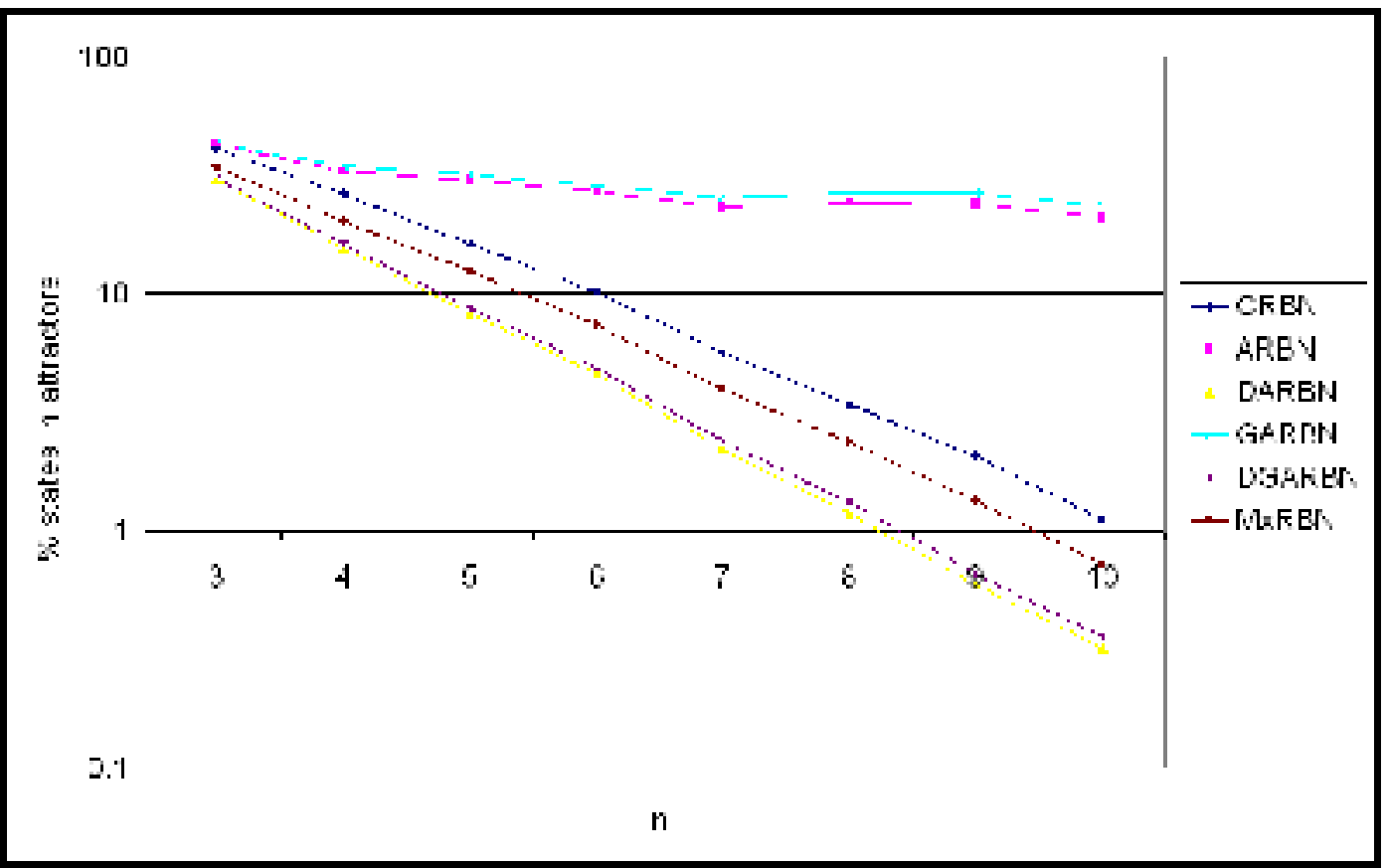}
\end{center}
\caption{Average percentage of states in attractors for k=3}
\label{pctk3}
\end{figure}

We can appreciate some of the results of our experiments in Figures \ref
{attk1}-\ref{pctk3} for $k=1$ and $k=3$. The averages are of one thousand
networks, exploring at most two thousand initial states. We used $maxP=7$
for contextual networks (DARBNs, DGARBNs, and MxRBNs). MxRBNs have $m=2$
pure contexts, randomly chosen at each $P=100$ time steps. The figures sum
point, cycle, and loose attractors.

The full data, including standard deviations, and more graphics ($k=2$) with
better resolution are available at http://homepages.vub.ac.be/\symbol{126}%
cgershen/rbn.

\subsection{Edge of chaos}

We have recently studied the sensitivity to initial conditions of different
types of RBNs \cite{GershensonSubmitted}. We first created randomly an
initial state A, and flip one node to have another initial state B. We run
each initial state in the network for ten thousand time steps, obtaining
states A' and B'. Then we compare the normalized Hamming distance (\ref
{hamming}) of the final states with the one of the initial states to obtain
a parameter $\delta $ (\ref{delta}).

\begin{equation}
H(A,B)=\frac{1}{n}\sum\limits_{i}^{n}\left\vert a_{i}-b_{i}\right\vert
\label{hamming}
\end{equation}

\begin{equation}
\delta=H_{t\rightarrow\infty}-H_{t=0}  \label{delta}
\end{equation}

If $\delta $ is negative, it means that the Hamming distance was reduced.
Since the initial distance is minimal ($\frac{1}{n}$), a negative $\delta $
indicates that both initial states tend to the same attractor. This implies
that the network is stable, in an ordered phase. A positive $\delta $
indicates that the dynamics for very similar initial states diverge. This is
a common characteristic of chaos in dynamical systems.

Since the initial states are chosen randomly, the comparison we make is
equivalent to see B as a perturbed version of A, and observe if the
perturbation affects the dynamics.

To compare the regimes of different types of RBN, we created $NN$ number of
networks (200), and evaluated for each $NS$ number of states (200) for all
six types of RBN.

\begin{figure}[t]
\begin{center}
\includegraphics[width=3.2in]{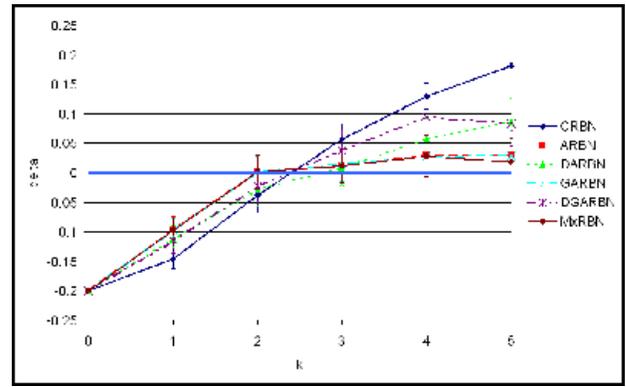}
\end{center}
\caption{Sensitivities to initial conditions for n=5}
\label{deltan5}
\end{figure}

We can observe the averages of $\delta $\ for networks with $n=5$ in Figure 
\ref{deltan5}. The error bars indicate the standard deviations. We can see
that all networks have an average phase transition from ordered to chaotic
for values of $k$ between one and three (although the standard deviations
indicate us that there can very well be chaotic networks for $k=2$ and
ordered for $k>2$). They have all a similar ``edge of chaos''.

Further results and analysis on the phase transitions of RBNs, including
statistics for networks of different sizes, can be found in \cite
{GershensonSubmitted}.

\section{Discussion}

We should say that all the presented data related to attractors have \emph{%
very} high standard deviations (this is why these are not shown in the
figures). This diversity is due to the fact that some networks have only one
attractor, others several. Some attractors are of length one (point), while
few ones are very long. This makes some networks to have very few states in
their attractors and others not so few. This high variance seems to be
common in these type of statistical studies \cite{MitchellEtAl1993}.

Nevertheless, one thing we can deduce from the first experiments, is that
all network types can perform \emph{complexity reduction}. This means that
compared to all the possible states of a network, only few ones are
favoured: those in attractors. This property has also been dubbed as ``order
for free'' \cite{Kauffman1993}. This is more noticeable as $k$ decreases:
fewer connections allow more order, and more order implies more complexity
reduction\footnote{%
But too much complexity reduction implies no adaptability nor evolvability.
That is why we find life at the ''edge of chaos''.}. Even when there is an
exponential increase of attractor length (states in an attractor) for ARBNs
and GARBNs, there is an even higher increase in complexity reduction, since
the percentage of states in attractors reduces exponentially with $n$ for
low values of k (see Figure \ref{pctk1}). ARBNs and GARBNs have also less,
but larger attractors, i.e. loose attractors include much more states. We
can see that the dynamics of non-deterministic RBNs with low $k$ are very
different from random transitions in a state space, because the percentage
of states in loose attractors is considerably small (independently of
standard deviations). We should notice that MxRBNs, even when
non-deterministic, reduce much more complexity than ARBNs and GARBNs, since
they have a constrained set of possible attractors, determined by their pure
contexts. This tells us that there are many ways in which a network can be
constrained in order to have complexity reduction. Other ways include
topological or functional bias. It is worth mentioning that this analysis
could not have been made without the inclusion of loose attractors in the
statistics. \qquad 

Another property independent of the updating scheme we have found is that
networks have almost the same phase transition from an ordered to a chaotic
phase. This is, they have very similar ``edge of chaos''. This region is
interesting, because it is precisely where computation and life can take
place \cite{Langton1990,FernandezSole2003}. It is flexible and stable enough
in order to allow the storage, transmission, and processing of information.
Thus, in principle, any updating scheme can be good for modelling life and
computation. Moreover, due to the high variances found, evolution has a
myriad of parameters to play with in order to find suitable networks. Not
only the size and connectivity are important. Also the functionality,
topology, updating scheme, and other criteria can be modified to reach an
``edge of chaos''. The precise numbers for this do not matter much for
theoretical studies, but we know that there is a much higher probability to
find it for values $1<k<3$ \cite{GershensonSubmitted}\footnote{%
In analytical studies $n$ has not been taken into account, but we have found
with simulations that the size of the networks does matter, since larger
nets seem to have indirect ``interferences'' which decrease the precise
critical value of $k$ as $n$ increases.} when there are no functional nor
topological biases. We speak about probabilities, but nature selects the
convenient configurations of parameters, making them more probable if they
are useful in a particular context. This changes the probability space and
the constraints of systems. Therefore, in some contexts we can expect to
find networks which reached their ``edge of chaos'' via functional bias,
others through topological bias, others through number of connections. But
in general, nature would fiddle with all of these parameters in order to
find viable networks.

However, it seems that determinism, or at least quasi-determinism, will be
favoured by evolution, since RBNs with these types of updating schemes can
perform more complexity reduction, as seen in Figure \ref{pctk3}. They can
have the similar stability than non-deterministic RBNs at the ``edge of
chaos'', but they will need, in general, less nodes to perform the same
functions. This can be thought in the following way: a non-deterministic
network can perform computations and complexity reduction, but in order to
cope with the non-determinism, a lot of redundancy is required. And
excessive redundancy costs. Contextual RBNs can perform even more complexity
reduction because they ``throw'' information into their context (but they
are more complex). We can assume that there are many constraints in nature
which prevent ``full'' non-determinism (where any node can be updated
indiscriminately), so in this case, plain ARBNs and GARBNs would not be good
models of natural networks. Still, they can be constrained through
artificial evolution to show rhythmic behaviour \cite{DiPaolo2001}. We can
also assume other constraints, such as delays \cite{KlemmBornholdt2003}, or
limited non-determinism, such as the one of MxRBNs.

Entering into the main debate concerning the use of synchronous RBNs as
realistic models, we have found that the main difference of ARBNs with
classical RBNs was due to non-determinism, not to asynchronicity \cite
{Gershenson2002e}. And the determinism of natural networks has just been
justified: deterministic or quasi-deterministic networks are more efficient,
especially in the case of large networks. There is not a big difference
between the properties of DARBNs and CRBNs, and moreover, they can be mapped
into one another \cite{Gershenson2002e}. Therefore, for \emph{simplicity},
CRBNs seem to be justifiable models of real networks, if we are interested
in theoretical studies of the possibilities of RBNs, such as the ones
carried out by Kauffman\footnote{%
Kauffman used RBNs to explain how there could be so few cell types with so
many genes. Indeed, there is a complexity reduction towards few cell types
(attractors). But the precise calculation related to the number of genes
(roughly known in 1993) and the precise number of expected cell types
(attractors) seems more like numerology, since there is a huge variance in
RBNs, and real regulatory networks do have topological and functional biases.%
}. This is because even when contextual RBNs can perform even more
complexity reduction\footnote{%
This is a bit tricky, because contextual RNBs have much more possible
states, since these include the actual phase of the context \cite
{GershensonEtAl2003a}.} than CRBNs, they are harder to study. CRBNs give
similar results than DARBNs and DGARBNs, and we should not be interested in
the \emph{precise} numbers we obtain. As we have seen, there is a huge
variance in RBNs, and different parameters, such as biases or constraints,
can change the precise numbers considerably. Another reason for justifying
CRBNs is that if we are not assuming any functionally, how could we assume
some updating period for a DARBN? Real networks are not fully synchronous,
but they are also not fully boolean, nor with homogeneous connectivity. We
believe that the synchronous assumption is justifiable for \emph{theoretical}
studies, especially compared to plain ARBNs. But if we are interested in
modelling a particular network, then the type of synchronicity should be
that which resembles more the one of the particular system modelled. Then we
should model a suitable updating scheme, but also a suitable functionality
and topology. DGARBNs seem to be a good alternative for this. MxRBNs are
also promising, since even when they are non-deterministic, they have enough
limits so that they can perform much better complexity reduction than plain
ARBNs or GARBNs. Other methods already mentioned are also worth exploring.

The main lesson from the presented data is that there is always a critical
region, and that nature \textit{thrives} (selects) towards it, since it is
of selective advantage. But in general, it does not matter which updating
scheme is being used, since all schemes \emph{have} this region. We can
conclude, as it has been stated by others, that life is very probable in our
universe \cite{Kauffman1993}, and almost inevitable in a planet like ours.
This contributes to the understanding of the general conditions for life.
But this understanding generates further questions. What about environments
which do not allow the exploration, selection, or retention of life? They
might be too ordered (frozen) or too chaotic (boiling). How abundant are
they? Which are the paths from them to a life-supportive environment? Which
are the paths \emph{out} of a life-supportive environment? Which are the
mechanisms used by an environment to \emph{maintain} or \emph{propagate} its
ability to promote life? These are questions which\ eagerly await
exploration.

\section{Conclusions}

In the XIX$^{th}$ century, many Latin American countries tried to develop
with ``order and progress'' (It was the lemma of Mexican president Porfirio
D\'{i}az, and the Brazilian flag bears the inscription ``Ordem E
Progresso''). In order to have both, a careful balance is required: too much
order does not allow changes, thus progress. Too much progress can
destabilize the order. The evolution of life requires the same ``order and
progress'' principle: order to retain acquired characteristics, progress to
explore new possibilities. This is the ``edge of chaos''. In this paper, we
have defended that in random Boolean networks, many parameters influence the
precise location of this region, but it exists, and evolution can find it.

Even when interactions in real systems may be non-deterministic, the
responses can be at a higher level deterministic, or close to deterministic.
It is convenient to have determinism, because of computational reasons:
information can be manipulated much easier and with less requirements. We
can see that for the same networks, deterministic updating offers much more
complexity reduction, therefore, this should be favoured by evolution. We
can assume that nature can find cyclic or quasi-cyclic behaviours, varying
different parameters, because it did, presumably more than once.

The main criticism to CRBNs was that the synchronicity was not justified 
\cite{HarveyBossomaier1997}. We believe that synchronicity can be justified
with our results, since we have seen that synchronous networks are able to
compute and to reduce complexity better than asynchronous non-deterministic
ones. Therefore, it is expected, and observed, that synchronicity will
evolve in living systems. How could this happen, and the precise mechanisms
by which asynchronous components can synchronize, are other questions, very
interesting ones, and people have been already studying them \cite
{RohlfshagenDiPaolo2004,Strogatz2003}.

\section{Acknowledgments}

I\ thank Jan Broekaert, Hughes Bersini, and two anonymous referees for
useful comments. This work was funded in part by the Consejo Nacional de
Ciencia y Tecnolog\'{i}a (CONACyT) of Mexico.

\bibliographystyle{alife9}
\bibliography{carlos,RBN}

\begin{thebibliography}{}

\bibitem[Aldana, 2003]{Aldana2003}
Aldana, M. (2003).
\newblock Boolean dynamics of networks with scale-free topology.
\newblock {\em Physica D}, 185(1).

\bibitem[Aldana-Gonz{\'a}lez et~al., 2003]{AldanaEtAl2003}
Aldana-Gonz{\'a}lez, M., Coppersmith, S., and Kadanoff, L.~P. (2003).
\newblock Boolean dynamics with random couplings.
\newblock In Kaplan, E., Marsden, J.~E., and Sreenivasan, K.~R., editors, {\em
  Perspectives and Problems in Nonlinear Science. A Celebratory Volume in Honor
  of Lawrence Sirovich}. Springer Applied Mathematical Sciences Series.

\bibitem[Bersini and Detours, 1994]{BersiniDetours1994}
Bersini, H. and Detours, V. (1994).
\newblock Asynchrony induces stability in cellular automata based models.
\newblock In {\em Proceedings of the {IVth} Conference on Artificial Life},
  pages 382--387. MIT Press.

\bibitem[Derrida and Pomeau, 1986]{DerridaPomeau1986}
Derrida, B. and Pomeau, Y. (1986).
\newblock Random networks of automata: A simple annealed approximation.
\newblock {\em Europhys. Lett.}, 1(2):45--49.

\bibitem[DiPaolo, 2001]{DiPaolo2001}
DiPaolo, E.~A. (2001).
\newblock Rhythmic and non-rhythmic attractors in asynchronous random boolean
  networks.
\newblock {\em Biosystems}, 59(3):185--195.

\bibitem[Fern{\'a}ndez and Sol{\'e}, 2003]{FernandezSole2003}
Fern{\'a}ndez, P. and Sol{\'e}, R. (2003).
\newblock The role of computation in complex regulatory networks.
\newblock Technical Report 03-10-055, Santa Fe Institute.

\bibitem[Gershenson, 2002]{Gershenson2002e}
Gershenson, C. (2002).
\newblock Classification of random boolean networks.
\newblock In Standish, R.~K., Bedau, M.~A., and Abbass, H.~A., editors, {\em
  Artificial Life {VIII}: Proceedings of the Eight International Conference on
  Artificial Life}, pages 1--8. MIT Press.

\bibitem[Gershenson, 2004]{GershensonSubmitted}
Gershenson, C. (2004).
\newblock Phase transitions in random boolean networks with different updating
  schemes.
\newblock {\em submitted}.

\bibitem[Gershenson et~al., 2003]{GershensonEtAl2003a}
Gershenson, C., Broekaert, J., and Aerts, D. (2003).
\newblock Contextual random boolean networks.
\newblock In Banzhaf, W., Christaller, T., Dittrich, P., Kim, J.~T., and
  Ziegler, J., editors, {\em Advances in Artificial Life, 7th European
  Conference, {ECAL} 2003 {LNAI} 2801}, pages 615--624. Springer-Verlag.

\bibitem[Harvey and Bossomaier, 1997]{HarveyBossomaier1997}
Harvey, I. and Bossomaier, T. (1997).
\newblock Time out of joint: Attractors in asynchronous random boolean
  networks.
\newblock In Husbands, P. and Harvey, I., editors, {\em Proceedings of the
  Fourth European Conference on Artificial Life {(ECAL97)}}, pages 67--75. MIT
  Press.

\bibitem[Kauffman, 1969]{Kauffman1969}
Kauffman, S.~A. (1969).
\newblock Metabolic stability and epigenesis in randomly constructed genetic
  nets.
\newblock {\em Journal of Theoretical Biology}, 22:437--467.

\bibitem[Kauffman, 1993]{Kauffman1993}
Kauffman, S.~A. (1993).
\newblock {\em The Origins of Order}.
\newblock Oxford University Press.

\bibitem[Klemm and Bornholdt, 2003]{KlemmBornholdt2003}
Klemm, K. and Bornholdt, S. (2003).
\newblock Robust gene regulation: Deterministic dynamics from asynchronous
  networks with delay.
\newblock q-bio/0309013.

\bibitem[Langton, 1990]{Langton1990}
Langton, C. (1990).
\newblock Computation at the edge of chaos: Phase transitions and emergent
  computation.
\newblock {\em Physica D}, 42:12--37.

\bibitem[Mitchell et~al., 1993]{MitchellEtAl1993}
Mitchell, M., Crutchfield, J.~P., and Hraber, P.~T. (1993).
\newblock Dynamics, computation, and the "edge of chaos": A re-examination.
\newblock Technical Report 93-06-040, Santa Fe Institute.

\bibitem[Rholfshagen and DiPaolo, 2004]{RohlfshagenDiPaolo2004}
Rholfshagen, P. and DiPaolo, E.~A. (2004).
\newblock The circular topology of rhythm in random asynchronous boolean
  networks.
\newblock {\em BioSystems}, 73(2):141--152.

\bibitem[Strogatz, 2003]{Strogatz2003}
Strogatz, S. (2003).
\newblock {\em Sync: The Emerging Science of Spontaneous Order}.
\newblock Hyperion.

\bibitem[Wuensche, 1997]{Wuensche1997}
Wuensche, A. (1997).
\newblock {\em Attractor Basins of Discrete Networks}.
\newblock PhD thesis, University of Sussex.

\end{thebibliography}

\end{document}